\documentclass[preprint,12pt,authoryear]{elsarticle}

\usepackage{appendix}

\usepackage{amssymb,amsmath}
\usepackage{amsthm}

\journal{arXiv}

\newcommand{\mR}{\ensuremath{{\mathbb R}}}
\newcommand{\mZ}{\ensuremath{{\mathbb Z}}}

\newcommand{\mE}{\ensuremath{{\mathbb E}}}

\newcommand{\ve}{\ensuremath{\varepsilon}}
\newcommand{\bA}{\underline{A}}
\newcommand{\Of}{\ensuremath{\mathcal O}_f}
\newcommand{\Ef}{\ensuremath{\mathcal E}_f}

\newcommand{\hOf}{\ensuremath{\hat{\Of}}}
\newcommand{\Kp}{\ensuremath{\mathcal K}_p}
\newcommand{\hKp}{\ensuremath{\hat{\Kp}}}
\newcommand{\Wf}{\ensuremath{\Xi_f}}

\newcommand{\hWf}{\ensuremath{\hat{\Wf}}}

\newcommand{\la}{\ensuremath{\langle}}
\newcommand{\ra}{\ensuremath{\rangle}}

\newcommand{\F}{\ensuremath{\mathcal F}}

\newcommand{\Pt}[1]{(#1_t)_{t \in \mZ}}

\theoremstyle{thmstyleone}%
\newtheorem{thm}{Theorem}
\newtheorem{Assumption}{Assumption}
\newtheorem{Lemma}{Lemma}
\newtheorem{corollary}{Corollary}

\newtheorem{Ex}{Example}

\begin{document}

\begin{frontmatter}


\title{Using Subspace Algorithms for the Estimation of Linear State Space Models for Over-Differenced Processes}


\author{Dietmar Bauer}

\affiliation{organization={Bielefeld University},
            addressline={Universitätsstrasse 25}, 
            city={Bielefeld},
            postcode={33615}, 
            state={NRW},
            country={Germany}}

\begin{abstract}
Subspace methods like {\em canonical variate analysis (CVA)} are regression
based methods for the estimation of linear dynamic state space models. They have been shown to deliver accurate (consistent and asymptotically equivalent to quasi maximum likelihood estimation using the Gaussian likelihood) estimators for invertible stationary autoregressive moving average (ARMA) processes. 
\\
These results use the assumption that the spectral density of the stationary process does not have zeros 
on the unit circle. This assumption is 
violated, for example, for over-differenced series that may arise in the setting of co-integrated 
processes made stationary by differencing. A second source of spectral zeros is 
inappropriate seasonal differencing to obtain seasonally adjusted data. This occurs, for example, by investigating yearly differences of processes that do not contain unit roots at all seasonal frequencies. 
\\
In this paper we show consistency for the CVA estimators for vector processes containing spectral zeros. 
The derived rates of convergence demonstrate that over-differencing can severely harm the 
asymptotic properties of the estimators making a case for working with unadjusted data.
\end{abstract}


\end{frontmatter}


\section{Introduction}
\label{sec:Intro}

Subspace algorithms such as the {\em Canonical Variate Analysis (CVA)} \citep{Larimore1983} are used for the estimation of linear dynamical state space systems for time series. 
CVA is popular since it is numerically cheap (consisting of a series of regressions), asymptotically equivalent to quasi maximum likelihood estimation (using the Gaussian likelihood) for stationary processes and robust to the existence of simple unit roots \citep[see][for a survey]{Bauer2005ESTIMATINGMETHODS}. This robustness also carries over to the case of seasonal unit roots, see \cite{BauerBuschmeier}. 

The algorithm fits a state space system in innovation form 
\begin{equation} \label{eq:statespace} 
y_t = Cx_t + \varepsilon_t, \quad x_{t+1} = A x_t + B \varepsilon_t, \quad t \in \mZ,
\end{equation} 
to an observed multivariate time series $y_t \in \mR^s, t= 1,...,T$. Here $A \in \mR^{n \times n}, B \in \mR^{n \times s}, C \in \mR^{s \times n}$ define the state space system with system order $n$, which must be supplied to the algorithm. In this paper we will always assume that the system is minimal \citep[implying that the state dimension cannot be reduced, see][Chapter 1, for details]{HanDei} and stable (such that all eigenvalues of $A$ are smaller than one in modulus).

The innovation form representation given above corresponds to the Wold representation of the stationary process $\Pt{y}$, if and only if the eigenvalues of the matrix $\bA =A-BC$ are inside or on the unit circle: In this case $\lambda_{|max|}(\bA) \le 1$ where $\lambda_{|max|}(M)$ denotes a maximum modulus eigenvalue of the matrix $M$. 

The asymptotic properties for CVA, when the data are generated from a stable state-space system, are documented in the literature; see, for example, \cite{Bauer2005ESTIMATINGMETHODS}.
However, results are restricted to the case of processes, where the strict inequality
$\lambda_{|max|}(\bA) < 1$ holds. 

This restriction may be violated, in particular, for
economic data, if the data are transformed to stationarity by
temporal differencing of all components. If co-integrating relations between the component variables exist and the whole time series is 
differenced, this leads to over-differencing in some 
directions introducing spectral zeros at frequency $\omega = 0$. Similar effects occur due to yearly differencing of time series, if not all unit roots to all seasonal frequencies are present. This happens, for example, if for an $I(1)$ process observed at quarterly frequency, yearly differences are examined. The corresponding seasonally differenced process then has spectral zeros for $\omega = \pi/2, \pi$ and $3\pi/2$.

In such a situation, the asymptotic properties of the subspace estimators currently are undocumented. This is matched by estimators obtained by maximizing the Gaussian likelihood. The results in \cite{HanDei}, for example, include systems with spectral zeros in the parameter set, but do not provide consistency for the transfer function estimators when the data are generated using such a system. Consequently, also no consistency rate is provided. In their Remark 1 on p. 126 consistency for $s=1$ is stated, but the general multivariate case is not dealt with. No result with regard to the asymptotic distribution is provided. \cite{Poetscher} investigated the asymptotic behaviour of maximum likelihood estimators in autoregressive moving average (ARMA) models for processes with spectral zeros and found severe problems with the likelihood maximizers in such situations. In some situations local maximizers in the vicnity of the data generating systems do not coincide with global optimizers. Moreover, since systems with spectral zeros lie at the boundary of the parameter region for typical ARMA parameterisations, standard asymptotic theory does not apply in this setting. 
Recently, \cite{Funovits} investigated estimators for non-invertible systems but again needed to exclude systems with zeros of the spectrum on the unit circle. Hence, currently, there is a gap in the literature with respect to the asymptotic properties of estimators in such situations.

This paper closes this gap to a certain extent for subspace procedures using results of \cite{Poskitt07} related to the autoregressive approximation of 
non-invertible processes. We show that CVA provides consistent estimators for the impulse response sequence also in 
the case of some spectral zeros. 
Consistency is obtained for the integer parameter $p$ of CVA (corresponding to the lag length of an autoregressive approximation) tending to infinity at a certain rate. We investigate the asymptotic bias arising for finite lag lengths and show that for typical choices it is not asymptotically negligible as it tends to zero slower than $1/\sqrt{T}$, the typical convergence rate involved in asymptotic normality. 

\section{Canonical variate analysis} \label{sec:CVA} 
The CVA method of estimation proposed by \citet{Larimore1983} is performed in three steps and uses two integers $f,p$
('future' and 'past') and information of the system order $n$  \citep[compare][]{Bauer2005ESTIMATINGMETHODS}:

\begin{enumerate}
    \item Obtain an estimate $\hat x_t \in \mR^n$ of the state $x_t$ for $t=p+1,...,T$.
    \item Estimate $C$ by regressing $y_t$ onto $\hat x_t$. This step provides residuals $\hat \ve_t = y_t - \hat C \hat x_t, t=p+1,...,T$.
    \item Estimate $A$ and $B$ by regressing $\hat x_{t+1}$ onto $\hat x_t$ and $\hat \ve_t, t=p+1,...,T-1$.
\end{enumerate}

The essential idea of CVA lies in the estimation of $x_t$ which uses the representation of the joint vector 
$Y_t^+ = (y_t',y_{t+1}',...,y_{t+f-1}')'$ for some integer $f\ge n$ as the state space system implies (using $E_t^+ = (\ve_t',\ve_{t+1}',...,\ve_{t+f-1}')'$)
\begin{equation} \label{eq:Hf}
Y_t^+ = \Of x_t + \Ef E_t^+, \quad x_t = \Kp Y_{t}^- + \delta x_t(p),
\end{equation} 
where $\Ef \in \mR^{fs \times fs}$ contains the impulse response coefficients and $\Of = (C',A'C',...,(A^{f-1})'C')' \in \mR^{fs\times n}$ denotes the observability matrix, which has full column rank due to minimality. 
Further, 

$$
\Kp  = \mE x_t (Y_t^-)' (\mE Y_t^- (Y_t^-)')^{-1}  \in \mR^{n \times ps}
$$
denotes the regression coefficient for explaining $x_t$ by $Y_{t}^- = (y_{t-1}',...,y_{t-p}')' \in \mR^{ps}$ for integer $p \ge n$ leading to the approximation $x_t(p)=\Kp Y_t^-$. Then $\delta x_t(p) = x_t - x_t(p)$ denotes the approximation error. 

As $x_t$ is not fully observed, $\Kp$ cannot be estimated directly. However, combining the two equations we obtain 

\begin{equation} \label{eq:CVA}
Y_t^+ = \Of \Kp Y_{t}^- + N_t^+ = \beta_{f,p} Y_t^- + N_t^+. 
\end{equation}

Here $N_t^+ = \Ef E_t^+ + \Of \delta x_t(p)$ is uncorrelated 
with $Y_t^-$ such that the OLS estimate $\hat \beta_{f,p} = \la Y_t^+, Y_t^- \ra \la Y_t^-, Y_t^-\ra^{-1}$ typically is consistent for fixed $f,p$. Here and below we use the notation 
$\la a_t, b_t \ra = T^{-1} \sum_{t=p+1}^T a_t b_t'$ for two processes $\Pt{a}$ and $\Pt{b}$. 

The matrix $\beta_{f,p} = \Of\Kp$ is of low 
rank $n \le \min(fs,ps)$ for $f,p$ chosen large enough. It 
follows that  estimates $\hOf, \hKp$ can be obtained using 
reduced rank regression techniques. Note, that such techniques 
also determine the split of the product $\hOf \hKp$ into 
factors $\hOf$ and $\hKp$ illustrating the identification 
issues in fixing the state basis.  
In order to identify the factors $\hOf$ and $\hKp$ from the product we use a selector matrix $S_f \in \mR^{n \times fs}$ such that $S_f \Of = I_n$. Such a matrix always exists \citep[cf., for example, the overlapping echelon forms, section 2.6 of][]{HanDei}. 
Then we impose the restriction $S_f \hOf = I_n$ to identify the system. 
For the results below corresponding to estimates of the impulse response coefficients (which are invariant in this respect) this choice of the state basis can be assumed without restriction of generality. 

The second and third step of CVA then amount to least squares using the estimate $\hat x_t = \hKp Y_t^-$ of the state:

\begin{eqnarray*}
\hat C  = \la y_{t}, \hat x_t \ra \la \hat x_t, \hat x_t \ra^{-1} & ,  & 
\hat \ve_t  =  y_t - \hat C \hat x_t, \\
\hat A  =  \la \hat x_{t+1}, \hat x_t \ra \la \hat x_t, \hat x_t \ra^{-1} & , & 
\hat B  = \la \hat x_{t+1}, \hat \ve_t \ra \la \hat \ve_t , \hat \ve_t \ra^{-1}.
\end{eqnarray*}

Note that these estimates contain two different sources of estimation error: (I) The deviation of sample moments from their population counter parts such as $\la y_t, y_{t-j} \ra - \mE y_t y_{t-j}'$ and (II) the approximation error of the state $\delta x_t(p)$. Here the first source contributes terms of order $O(\sqrt{(\log T)/T})$ typically (see below).\footnote{In fact often as slightly smaller upper bound $\sqrt{\log \log T/T}$ is obtained from the law of the iterated logarithm. The difference is due to the required uniformity in the lag length for a wide range of lag lengths.}  
With respect to (II) under the strict minimum-phase assumption implying $\bA^p \to 0$ for $p \to \infty$ 
we may use
$\tilde \Kp = [B, \bA B, \bA^2 B ,..., \bA^{p-1}B]$
leading to $x_t = \tilde \Kp Y_t^- + \bA^{p}x_{t-p}$
to infer that the variance of the approximation error $\delta x_t(p)$  can be bounded by 
$\bA^p (\mE x_{t-p} x_{t-p}') (\bA^p)'$ such that it is of order $O(\rho_\circ^{2p})$ where $1>\rho_\circ > \lambda_{|max|}(\bA)$: If in that case $p = p(T) = -\frac{(1+\epsilon)(\log T)}{2\log \rho_\circ}$ (or rather its integer part) is used, we obtain 

$$\rho_\circ^p = exp(-\frac{(1+\epsilon)(\log T)}{2\log \rho_\circ} \log \rho_\circ)= exp(-\frac{(1+\epsilon)}{2}\log T) = T^{-(1+\epsilon)/2}
$$

\noindent such that the variance of the approximation error is of order $T^{-(1+\epsilon)}$. If $\epsilon>0$ this implies that the approximation error is negligible in the usual $\sqrt{T}$ asymptotics. 

For $\rho_\circ=1$ this argument does not work any more. \cite{Poskitt07} shows that also in this  'non-invertible' case the approximation error decreases to zero albeit not at the same speed.\footnote{As \citet{Funovits} points out, the term 'non-invertible' for this situation is inaccurate, as the system may be inverted, but not with the usual tools. This is demonstrated in Example~\ref{ex:one}. } 

\begin{Ex} \label{ex:one}
Consider $y_t = \ve_t - \ve_{t-1} \in \mR^s$ for independent identically distributed white noise $\Pt{\ve}$ with expectation zero and variance $\Omega>0$. This can be represented in state space form as 

$$
y_t = I_s x_t + \ve_t, \quad x_{t+1} = 0_{s \times s}  x_t - I_s \ve_t
$$

\noindent and hence $x_t = -\ve_{t-1}$ and $(A,B,C) = (0_{s \times s},-I_s,I_s)$. Following \cite{Poskitt07} we see that 
$\Kp  =  -[\frac{p}{p+1}I_s,\frac{p-1}{p+1}I_s,...,\frac{1}{p+1}I_s]
$
implying that 
\begin{eqnarray*}
x_t(p) 
& = & \Kp Y_t^-  = \frac{-1}{p+1}\sum_{j=1}^p (p+1-j) (\ve_{t-j} - \ve_{t-j-1}) \\
& = & \frac{-1}{p+1}\sum_{j=1}^p (p+1-j) \ve_{t-j} - \frac{-1}{p+1}\sum_{j=2}^{p+1} (p+2-j) \ve_{t-j}\\
& = & \frac{-1}{p+1}  \left( p \ve_{t-1} - \sum_{j=2}^p \ve_{t-j}  - \ve_{t-p-1}\right) 
= -\ve_{t-1} + \frac{1}{p+1}\sum_{j=1}^{p+1} \ve_{t-j}.
\end{eqnarray*} 

\noindent Denoting $\overline{\ve}_{t-1}(p) = \sum_{j=1}^{p+1} \ve_{t-j}/(p+1)$ 
we obtain $x_t = x_t(p) - \overline{\ve}_{t-1}(p), \ve_t(p)= y_t - x_t(p) = \ve_t - \overline{\ve}_{t-1}(p)$
such that the approximation error $\delta x_t(p) = -\overline{\ve}_{t-1}(p)$. 
It follows that $\mE \delta x_t(p) \delta x_t(p)' = \frac{1}{p+1}\Omega$.
Thus the approximation error tends to zero in mean square, but the variance is of order $1/p$ and not $\rho_\circ^{2p}$ decreasing much slower as a function of the sample size for $p=p(T)$. \qed 
\end{Ex}

This example is typical. The same arguments show that the variance of the approximation error for $y_t = \Delta u_t = u_t - u_{t-1}$ for stationary process $\Pt{u}$ with non-singular spectral density at $\omega = 0$ (not necessarily white noise) is at most of order $p^{-1}$. 

\section{Consistency of the Estimates} \label{sec:res} 
\cite{Poskitt07} derives results for the estimation accuracy for the autoregressive approximation coefficients: In his Theorem 5 he states that uniformly in $0 < p \le H_T$
for some upper bound $H_T = O(\sqrt{T/\log T})$ and using $Q_T^2 = (\log T)/T$ we have ($\lambda_{min}(M)$ denoting the smallest eigenvalue of the symmetric matrix $M$)

\begin{equation} \label{eq:bound}
\sum_{j=1}^p | \hat \alpha_p(j) - \alpha_p(j)|^2 = O\left( 
\frac{p}{\lambda_{min}(\Gamma_p)^2} Q_T^2 \right)
\end{equation} 

\noindent where $O(.)$ denotes almost sure convergence at the given rate. Here $\alpha_p(j), j=1,...,p,$ denote the autoregressive coefficients in a lag $p$ approximation for $\Pt{y}$ obtained from 

$$
[\alpha_p(1),...,\alpha_p(p)] = \left( \mE y_t (Y_t^-)' \right)\Gamma_p^{-1}, \quad \Gamma_p =  \mE Y_t^- (Y_t^-)'
$$

\noindent and $\hat \alpha_p(j)$ are the corresponding least squares estimates. 
\cite{Poskitt07} uses a univariate setting, however, the extension to multivariate time series in our framework is straightforward taking the lower bound on the eigenvalues of $\Gamma_p$ as given in Lemma~\ref{lem:bound} below into account. 

In this paper we do not investigate autoregressive processes but state space processes with spectral zeros. We focus on the case of simple spectral zeros obtained by one time over-differencing:

\begin{Assumption} \label{ass:dgp} 
The stationary process $\Pt{y}, y_t \in \mR^s,$ is generated using a rational, stable and invertible transfer function $\tilde k(z)= I_s + \sum_{j=1}^\infty  \tilde K_j z^j, \tilde K_j = \tilde C_\circ \tilde A_\circ^{j-1}\tilde B_\circ$ (which hence has all its zeros and poles outside the unit circle) where $\tilde A_\circ \in \mR^{\tilde n \times \tilde n}$ and an orthonormal matrix $M = [M_c,M_{s-c}] \in \mR^{s \times s}, M'M=I_s, M_c \in \mR^{s \times c}$, where $0<c \le s$ is an integer, as ($L$ denoting the backward-shift operator and $\Delta = (1-L)$)

$$
\Pt{y}  = M \left[ \begin{array}{cc} \Delta I_c & 0 \\ 0 & I_{s-c} \end{array} \right] M' \tilde k(L) \Pt{\ve}
= k(L) \Pt{\ve}.
$$
\noindent The transfer function 

$$
k(z)
= M \left[ \begin{array}{cc} \Delta I_c & 0 \\ 0 & I_{s-c} \end{array} \right] M' \tilde k(z) 
= I_s + \sum_{j=1}^\infty C_\circ A_\circ^{j-1} B_\circ z^j
$$

is represented as 
$(A_\circ,B_\circ,C_\circ), A_\circ \in \mR^{n\times n}, n= \tilde n+c$, where 
the corresponding observability matrix 

$$
\Of := \begin{pmatrix}
C_\circ \\ C_\circ A_\circ \\ \vdots \\
C_\circ A_\circ^{f-1}
\end{pmatrix}
$$

fulfills the restriction $S_f \Of = I_{n}$ for selector matrix $S_f \in \mR^{n \times fs}$. 
\\
\noindent Here $\Pt{\ve}$ denotes a zero mean ergodic, stationary, martingale difference sequence with respect to the sequence $\F_t$ of sigma-fields spanned by the past of $\ve_t$ fulfilling 

\begin{eqnarray*}
\mE ( \ve_t | \F_{t-1} ) = 0 & , & \mE ( \ve_t \ve_t' | \F_{t-1}) =\mE ( \ve_t \ve_t') =  \Omega. 
\end{eqnarray*}

\noindent Furthermore $\mE \ve_{t,j}^4 < \infty, j=1,...,s$. 
\end{Assumption} 

We use the same noise assumptions as \cite{Poskitt07} and \cite{HanDei}. Clearly such processes have a spectral density of rank $s-c$ (which is hence singular) for $\omega =0$ due to the differencing. At all other frequencies, the rank 
equals $s$ since $\tilde k(L)$ is assumed to be invertible. 

Note that under these assumptions, we have 
\begin {align*}
k(z) &= 
I_s + z \begin{pmatrix} \tilde C_\circ, & - M_c \end{pmatrix} \left( I_n - z 
\begin{pmatrix} \tilde A_ \circ & 0 \\ M_c'\tilde C_\circ & 0  \end{pmatrix}\right)^{-1} \begin{pmatrix}
   \tilde B_\circ \\ M_c'  \end{pmatrix}. 
\end{align*}

This representation not necessarily is minimal, and does not necessarily fulfill $S_f \Of=I_{n}$. This implies that the process $\Pt{y}$ is generated by a state space system which is stable, but not strictly minimum-phase. 

Such a representation is obtained, for example, when examining first differences of an I(1) autoregressive moving average process generated from a state space system.  
Using the vector of seasons representation we obtain a similar 
representation for yearly differences of processes that are 
integrated at other seasonal frequencies:  \citet{Bauer2019} 
demonstrates that the vector of seasons representation of such 
processes is an $I(1)$ process.   

The bound in~\eqref{eq:bound} contains $\lambda_{min}(\Gamma_p)$ which depends on the data generating process. 
A multivariate extension to Theorem 2 of  \cite{Palma2003} provides a characterization of  $\lambda_{min}(\Gamma_p)$ (the proof of the lemma is given in the Appendix):

\begin{Lemma} \label{lem:bound}
Let the process $\Pt{y}$ be generated according to Assumptions~\ref{ass:dgp}. Define $\Gamma_p = \mE Y_t^- (Y_t^-)'$ for $Y_t^- = (y_{t-1}',...,y_{t-p}') \in \mR^{ps}$. 
\\
Then $\lambda_{min}(\Gamma_p)^{-1} = O(p^2)$ as a function of $p \to \infty$. 
\end{Lemma}

This implies that the bound in~\eqref{eq:bound} amounts to $p(T)^5 \log T/T$ which tends to zero, if $p(T) = c \lfloor T^{\delta} \rfloor$ for $0< \delta < 0.2$. Note, however, that for this rate of increase the approximation error $\delta x_t(p)$ (with variance of order $p(T)^{-1}$, see above) is larger than $O(1/\sqrt{T})$ and hence dominates the asymptotic distribution of terms like $\sqrt{T}(\hat A- A_\circ)$. 

Due to the structure of the CVA algorithm, the results from the autoregressive setting can be used almost immediately 
for the CVA setting, if $f \ge n$ fixed and $p = f\tilde p$ where $\tilde p = \tilde p(T) = o(T^{\delta})$ depends on the sample size. 
This implies that for the approximation of $x_t$ the unrestricted estimate 
$\hat \beta_{f,p} = \la Y_t^+ , Y_t^- \ra \la Y_t^- , Y_t^- \ra^{-1}$ equals an autoregressive model for $Y_t^+$. 
This matrix -- which in the limit has rank $n$ -- then is low rank approximated leading to the estimate $\hOf\hKp$ of $\Of \Kp$. Low rank approximations typically  retain the error bound (see the proof of Theorem~\ref{thm:cons} below), such that 
$\| \hOf \hKp - \Of \Kp \| = O( \| \hat \beta_{f,p} - \Of\Kp \|)$ for fixed $f,p$. 


The second and third step of CVA then amount to least squares using the estimate $\hat x_t = \hKp Y_t^-$ of the state. If instead we had access to the state approximation $x_t(p) = \Kp Y_t^-$ as well
as population instead of sample moments 
we would obtain the following quantities: 

%
%
\begin{eqnarray*}
C_p  =  \mE y_{t}x_t(p)' (\mE x_t(p) x_t(p)')^{-1} &,& \ve_t(p) = y_t - C_p x_t(p), \\
A_p  =  \mE x_{t+1}(p)x_t(p)' (\mE x_t(p) x_t(p)')^{-1} &,& 
B_p  =  \mE x_{t+1}(p)\ve_t(p)' (\mE \ve_t(p)\ve_t(p)')^{-1}.
\end{eqnarray*}

\noindent 
Here we emphasize in the notation the dependence on the approximation lag length $p$. 
If the approximation errors tend to zero and the convergence of sample covariances to population quantities is uniform in $p$ then consistency for $p \to \infty$  follows (for the proof see the Appendix): 

\begin{thm} \label{thm:cons} 
Let the process $\Pt{y}$ be generated according to Assumptions~\ref{ass:dgp} where $c>0$. Let the CVA procedure be applied with $f \ge n$ not depending on $T$ and $p = p(T) \to \infty$ for $T \to \infty$ such that $p(T) = o(T^{\delta}), 0 < \delta < 0.2$. 

Then: 
\begin{eqnarray*}
\max \{ \| \hat A - A_p \|, \| \hat B - B_p \|, \| \hat C - C_p \| \} & = & O(\sqrt{p^{5}(\log T)/T}), \\
\max \{ \| A_\circ -  A_p \|, \| B_\circ -  B_p \|, \| C_\circ -  C_p \| \} & \to & 0
\end{eqnarray*}

\noindent for $p=p(T) \to \infty$ as $T \to \infty$.
Consequently $\hat C\hat A^j \hat B \to C_\circ A_\circ^j B_\circ, j=0,1,2,...$ almost surely.
\end{thm}

Here the convergence of the system matrix estimators uses the normalization $S_f \Of=I_n$ and similarly for the estimated system. This is only possible, if $S_f \Of$ is 
non-singular. Using the overlapping  echelon forms \citep[see][ chapter 2]{HanDei} this holds generically in the set of all transfer functions of order $n$. For every transfer function such a choice exists. Thus, the particular choice is not seen as critical. For the convergence of the impulse response sequence, knowledge of $S_f$ is not necessary.

Note that these two error bounds are differently influenced by the integer $p$: large $p$ reduces the approximation errors such as $A_p-A_\circ$ but increases the sampling error $\hat A- A_p$. Both tend to zero slower for spectral zeros than under the strict minimum-phase assumption. 

\addtocounter{Ex}{-1}
\begin{Ex}[{\bf continuation}] \label{ex:bias} 
Consider again $y_t = \Delta \ve_t$ for white noise $\Pt{\ve}$. Then 
$x_t(p) =  -\ve_{t-1} +\overline{\ve}_{t-1}(p)$ and $\ve_t(p)= \ve_t - \overline{\ve}_{t-1}(p)$.
It follows that $\mE x_t(p)\ve_t(p)' = 0,$
\begin{eqnarray*}
\mE \ve_t(p)\ve_t(p)' =  \frac{p+2}{p+1}\Omega & , & \mE x_t(p) x_t(p)'  = \frac{p}{p+1}\Omega, \\
\mE x_{t+1}(p) \ve_t(p)' = -(1-\frac{1}{(p+1)^2}) \Omega & , & 
\mE x_{t+1}(p) x_t(p)' = - \frac{1}{(p+1)^2} \Omega.
\end{eqnarray*}
Thus  $A_p = A_\circ -I_s \frac{1}{p(p+1)}, B_p =  B_\circ + \frac{1}{p+1} I_s, C_p = C_\circ$.  
\end{Ex}

This shows for the special case that the system $(A_p,B_p,C_p)$ for fixed $p$ is a biased estimate of the true system $(0,-I_s,I_s)$. The bias is of order $p^{-1}$. In 
order for this bias to be asymptotically negligible $p$ has to grow 
faster than $T^{1/2}$. This is faster than the upper bound $T^\delta$ or even the bound $H_T = \sqrt{T/\log T}$ used above, such that with our methods we cannot 
derive results for the asymptotic distribution of the system 
estimates. Note, that in this situation the smallest eigenvalue of $\Gamma_p$ tends to zero as $p^{-2}$ which then is of order $O(1/T)$. This implies that the inverse $\Gamma_p^{-1}$ amplifies the sampling error in some directions as $O(T)$, which is larger than the estimation precision for sample second moments. 

Additionally note that typically the upper bound for selecting the lag length is $H_T = c \lfloor T^{1/4} \rfloor$ such that the bias derived above is the dominant term in the asymptotics. 

Similar biases are expected in the general case different from $k(z) = \Delta I_s$ as used in Example~1, as it is the approximation of the inverse of $\Delta$ that introduces the issues. 

Note that this contrasts the case $c=0$ without over-differencing, where we obtain for the choice 
$p(T) \ge -(1+\epsilon)\log T/(2\log \rho_\circ), p=O((\log T)^a)$ for some $\epsilon>0$ \citep[see, for example, the survey][]{Bauer2005ESTIMATINGMETHODS}

\begin{align*}
\max \{ \|  A_\circ -  A_p \|, \|  B_\circ -  B_p \|, \|  C_\circ -  C_p \| \} &=o(1/\sqrt{T}), \\
\max \{ \|  A_\circ -  \hat A \|, \|  B_\circ -  \hat B \|, \|  C_\circ -  \hat C \| \} &= O(\sqrt{(\log \log T)/T}).
\end{align*}

After examining the bias term we also provide a result on the asymptotic distribution of the estimators. The result is only indicative, as it uses the assumption of a fixed lag length $p$ not depending on $T$. We do not attempt to derive a result uniform in $p$ although this is likely to hold. The technical complication is considerable and the potential gain is small given that the bias dominates the variability of the estimator. 

\begin{thm} \label{thm:asynorm}
Let the process $\Pt{y}$ be generated according to Assumptions~\ref{ass:dgp} where 
$c>0$. Let the CVA procedure be applied with $f \ge n$ not depending on $T$ and $p \ge n$ not depending on $T \to \infty$. 
\\
Then for each $j = 1,2,...$ we have
\begin{eqnarray*}
\sqrt{T} \mbox{vec} \left( \hat C \hat A^{j-1} \hat B - C_p A_p^{j-1}B_p \right) \stackrel{d}{\to} {\mathcal N}(0, V_p).
\end{eqnarray*}
\end{thm}
The theorem shows that the estimators of the impulse response sequence are asymptotically normally distributed with the usual $\sqrt{T}$ rate around the impulse response $C_pA_p^{j-1}B_p$ corresponding to the system $(A_p,B_p,C_p)$. The proof is found in the Appendix. 

Finally, the results can easily be generalized to the case of yearly differencing seasonally integrated processes: 

\begin{corollary}
Let $\Pt{y}$ be a multi-frequency I(1) process in the sense of \cite{BauWag} observed with a frequency of $S$ observations per year, such that the  yearly differences $\Pt{w}=(1-L^S)\Pt{y}$ constitute a stationary process with the representation 
$$
\Pt{w} = k(L)\Pt{\ve}
$$
where $\Pt{\ve}$ is as in Assumption~\ref{ass:dgp} and $k(z)=I_s + zC_\circ(I_n-zA_\circ)^{-1}B_\circ$ where 
$A_\circ$ is stable and $\det k(z) \ne 0, |z| \le 1$ with the possible exception of $z_k = \exp(i 2\pi k/S),k=0,...,S-1$. 
\\
Then the impulse response estimates $\hat C\hat A^{j-1}\hat B$ obtained from CVA with $f \ge n, p=p(T) \to \infty$ such that $p(T)=o(T^\delta), 0 < \delta < 0.2$  are consistent.  
\end{corollary}

The proof follows from noting that the vector of seasons 
representation for an MFI(1) process with unit root 
frequencies being equal to the seasonal frequencies $\omega_k = 2\pi k/S$ converts the process to an I(1) process. Then the
theorem above can be applied, noting that we can always resort
to subvectors.

\section{Simulation} \label{sec:simul}

In this section we simulate a test example to indicate the relative performance of three different estimators for an over-differenced time series:

\begin{itemize}
    \item CVA: the subspace procedure described above
    \item qMLE: maximum likelihood estimation based on the Gaussian likelihood. Here both stability and minimum-phase assumption are imposed using a barrier function approach.
    \item PEM: prediction error methods use the assumption of $x_0=0$. With this intialisation the Kalman filter collapses to the inverse system. Again stability is enforced using a barrier approach. The minimum-phase assumption, however, is not imposed. For systems that are not minimum-phase the Kalman filter is unstable which introduces a penalisation for such systems. 
\end{itemize}

qMLE and PEM are initialised using the CVA estimate. As the data generating process we use a bivariate system:

\begin{eqnarray*}
y_t  & = & \begin{pmatrix} 0.7 & 0 \\ 0 & 0.2 \end{pmatrix} x_t + e_t, \\
x_{t+1} & = & \begin{pmatrix} 0.7 & 0 \\ 0 & 0.2 \end{pmatrix} x_t + \begin{pmatrix}
   1 & 0  \\ 0 & 1  \end{pmatrix} e_t.
\end{eqnarray*}

Here $\Pt{e}$ denotes a bivariate standard normal error process. Consequently, the 
process is a bivariate AR(1) process with independent innovations, where the state equals $y_{t-1}$. We apply the 
estimation procedures to $\Delta \Pt{y}$, which has the state space representation 

\begin{eqnarray*}
\Delta y_t  & = & \begin{pmatrix} -0.3 & 0 \\ 0 & -0.8 \end{pmatrix} x_t + e_t, \\
x_{t+1} & = & \begin{pmatrix} 0.7 & 0 \\ 0 & 0.2 \end{pmatrix} x_t + \begin{pmatrix}
   1 & 0  \\ 0 & 1  \end{pmatrix} e_t.
\end{eqnarray*}

We generate $M=1000$ data sets of sample size $T \in \{ 100,200,400,800,1600 \}$
and estimate a state space system with $n=2$ for each data set. Hereby $\hat f= \hat p = 2\hat k_{AIC}$ is chosen, where $\hat k_{AIC}$ denotes the lag length of an autoregressive approximation chosen using {\tt AIC}. 

\begin{figure}  
\begin{tabular}{ccc}
\includegraphics[width=4.5cm]{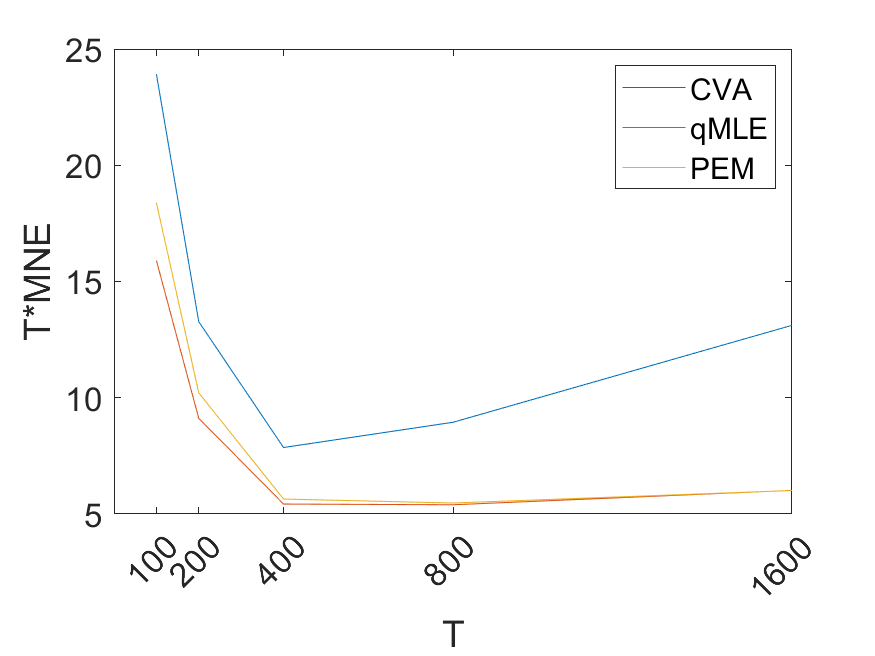} & 
\includegraphics[width=4.5cm]{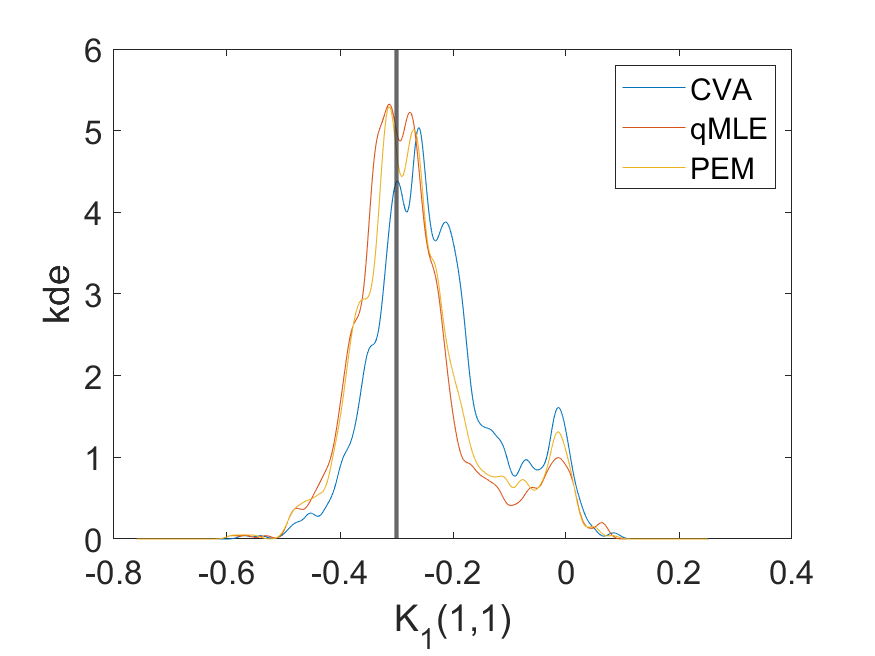} & 
\includegraphics[width=4.5cm]{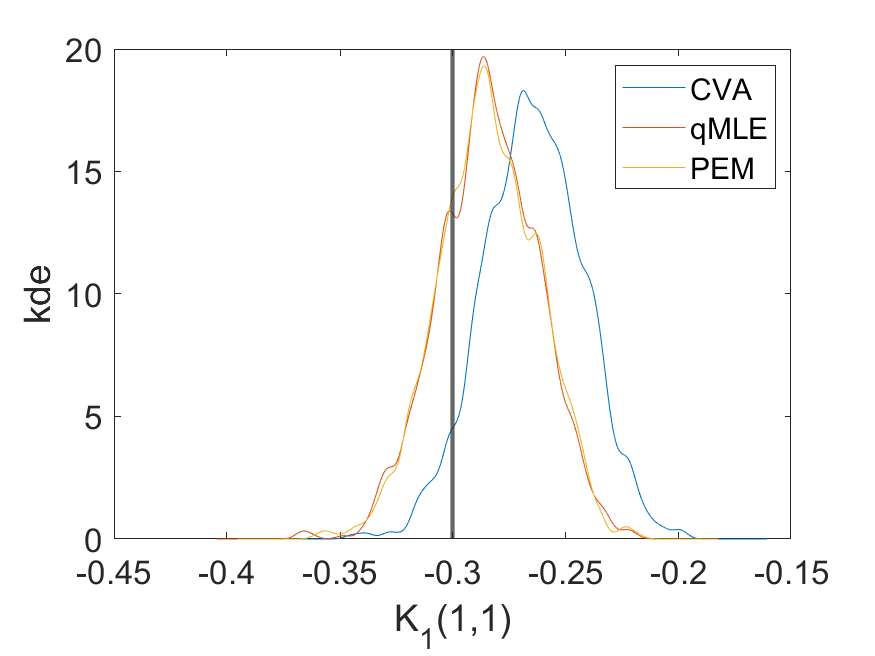} \\
(a) & (b) & (c) 
\end{tabular} 
\caption{(a) Mean squared norm of impulse response estimation error times sample size. (b) Density estimate of estimates of $K_1(1,1)$ for $T=200$. (c) Density estimate of estimates of $K_1(1,1)$ for $T=1600$. } \label{fig:approx}
\end{figure}

The results can be seen in Figure~\ref{fig:approx}. Plot (a) of that figure 
provides a plot of the mean squared error of the impulse response estimates times 
the sample size $T$. A convergence rate of order $O_P(T^{-1/2})$ would imply that the curves level off for 
large sample sizes. While this seems to be the case for qMLE and PEM, the curve for {\tt CVA} shows an increase for the larger sample size. The decrease at the start 
is due to the decrease in variance, while for the largest sample size $T=1600$ the 
pronounced bias in the {\tt CVA} estimate leads to an increase in the variance of 
the normalized impusle response estimate. 

\begin{figure}  
\begin{tabular}{ccc}
\includegraphics[width=7cm]{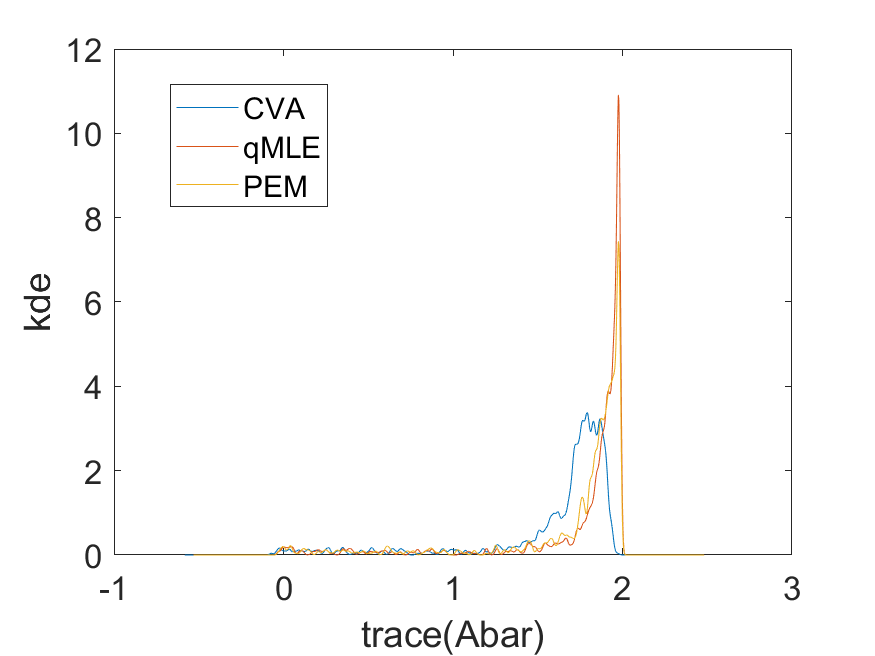} & 
\includegraphics[width=7cm]{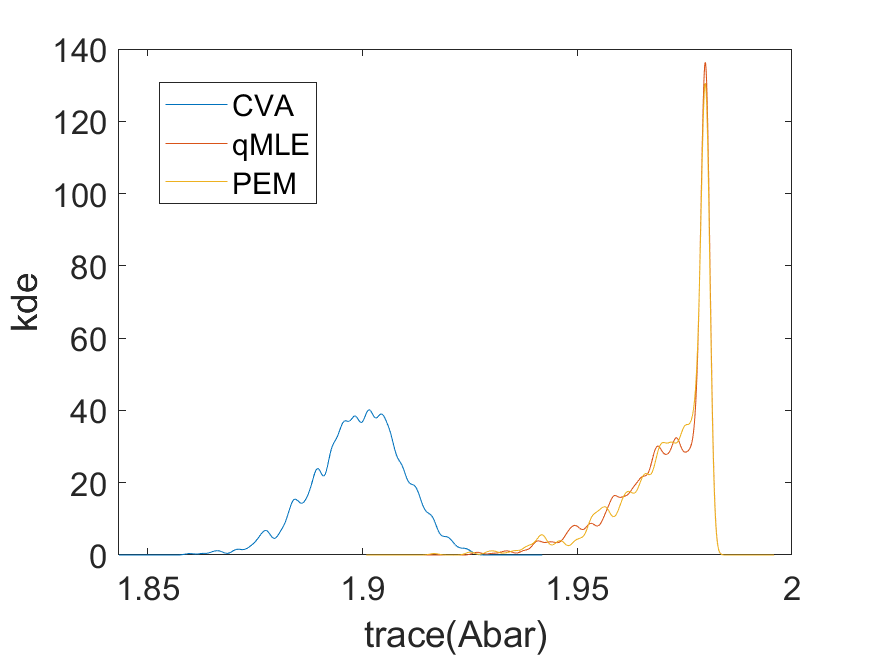} \\
(a) & (b) \\ 
\includegraphics[width=7cm]{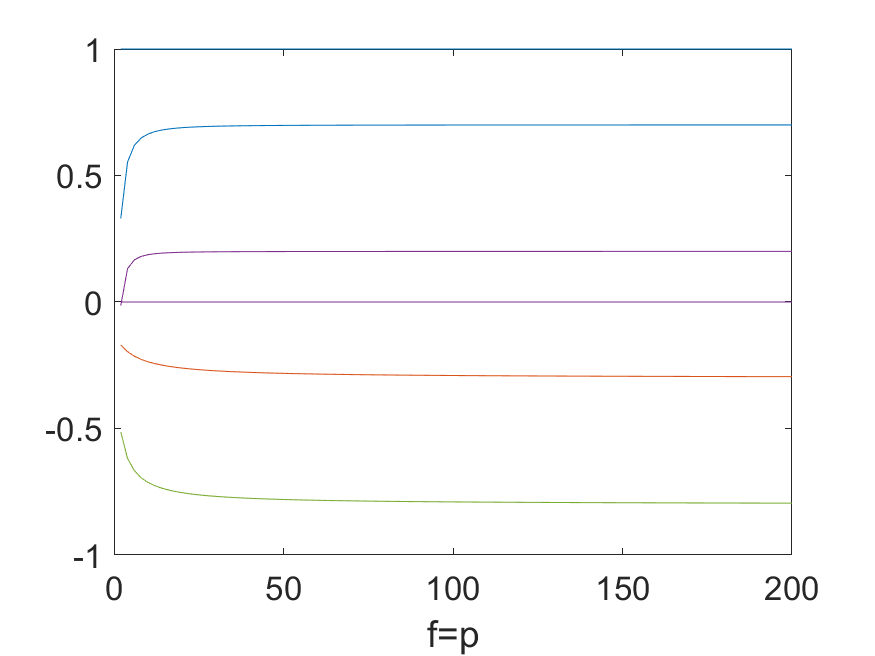} &
\includegraphics[width=7cm]{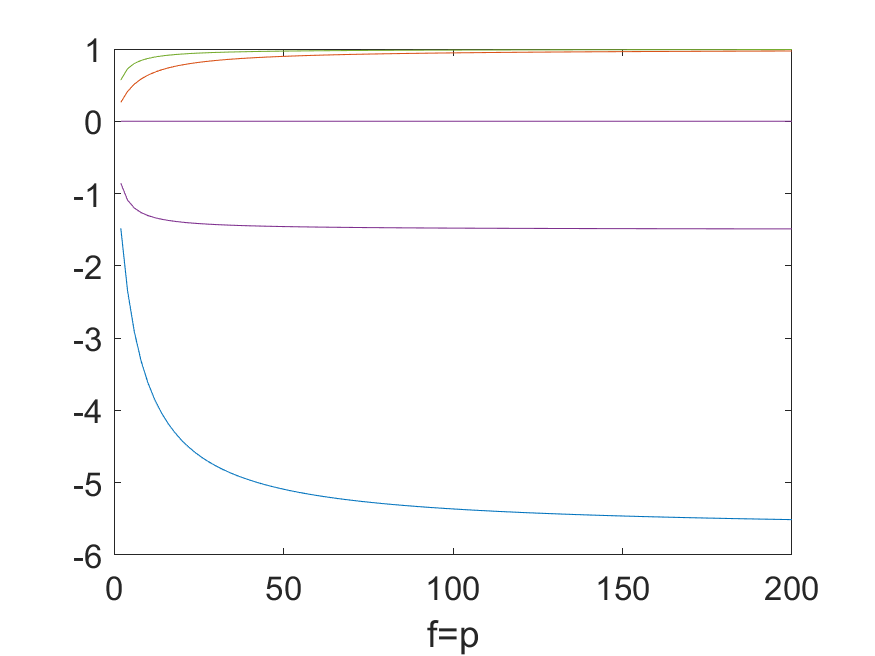} \\
 (c) & (d) 
\end{tabular} 
\caption{Density estimate of the estimated trace of $\underline{A}$ for $T=200$ (a) and $T=1600$ (b). For numerical reasons the barrier is located for eigenvalue equal to 0.99. (c): Dependence of $(A_p,B_p,C_p)$ on $p$. (d): Scaled deviations of $(A_p,B_p,C_p)$ from $(A_\circ,B_\circ,C_\circ)$ as a function of $p$. } \label{fig:bias}
\end{figure}

While the impulse response estimates for the large sample size appear to be Gaussian distributed, this is not the case for all system dependent quantities. Figure~\ref{fig:bias} (a) and (b) provides density estimates for the estimate of the trace of $\underline{A} = A-BC$. 
For $\underline{A}_\circ = A_\circ-B_\circ C_\circ = I_2 \in \mR^2$ the trace equals $2$. The minimum-phase assumption implies that all eigenvalues need to be smaller than 1 in modulus. Hence $\mbox{tr}(\underline{A}) \le 2$ must hold for all estimates from qMLE and is likely to hold for PEM estimates. This is visible for both sample sizes. The trace for qMLE and PEM clearly is not Gaussian distributed with a strong discrete component at $2$. For {\tt CVA} the situation is different: The bias results in smaller values at both sample sizes and the estimates appear to be well represented as a normal distribution around the biased value.

The lower row of plots in Figure~\ref{fig:bias} demonstrates the size of the bias in the {\tt CVA} quantities $(A_p,B_p,C_p)$ calculated using the {\tt CVA} algorithm using the true covariances in place of their estimates. 
The evaluations in the example suggest that the bias in $A_p$ is of order $O(p^{-2})$, while the bias in the entries of $B_p$ is of order $O(p^{-1})$. 
Figure~\ref{fig:bias} (c) provides a plot of the entries in the system as a function of $f=p$, while (d) scales by multiplying the deviations $A_p-A_\circ$ by $p^2$ and the deviations in the remaining system matrices by $p$. 
We observe the decrease in the bias, which according to the scaled plot (d) is of 
the order given in the example.

\section{Conclusions} \label{sec:concl}
In this paper we show that working with first differences does not invalidate consistency for CVA. This is a relief in situations where one is not sure about the existence of cointegrating relations or the presence of all seasonal unit roots, when working with (seasonal) differences rather than the original measurements. 

Inference, on the other hand, gets more complicated as the asymptotic distribution in a situation, where some of the variables are over-differenced, is not known, contrary to the case of no over-differencing. Our results show that in this situation the estimates suffer from a relatively large bias term preventing the usual $\sqrt{T}$ asymptotic normality and hence inference for subspace estimates is non-standard in these cases. 

The results imply that also higher order of differencing  
can be dealt with using exactly the same methods, but making things even worse. In such 
situations consistency of CVA estimators of the impulse 
response sequence again follows for $p$ increasing 
sufficiently slow. 

However, in all these cases the rate of consistency is slower than $1/\sqrt{T}$. Hence the main message of this paper is to prefer the original, un-differenced time series for inference.

\section*{Declarations}

\begin{itemize}
\item Funding: This research was funded by the Deutsche Forschungsgemeinschaft (DFG, German Research Foundation - Projektnummer 469278259) which is gratefully acknowledged.
\item Conflict of interest/Competing interests: The author declares that there are no competing interests.
\end{itemize}

\begin{appendices}

\section*{Proof of Lemma~\ref{lem:bound}}

Theorem 2 of \cite{Palma2003} deals with the univariate case 
and provides the lower bound for the eigenvalues of 
$\Gamma_p$. The spectral zeros in that theorem can be at arbitrary locations with different multiplicity. The order of the smallest eigenvalue then depends on the largest multiplicity. In the current case we only have simple eigenvalues such that $d_1=-1$ there. 
\\
In our situation we have 

\begin{align*}
f(\omega)  &=  \frac{1}{2\pi}   k(e^{i\omega})\Sigma k(e^{i\omega})^* \\
&\ge M \begin{pmatrix}(1-e^{i\omega}) I_{c} & 0\\ 0 & I_{s-c} \end{pmatrix} \begin{pmatrix}(1-e^{-i\omega}) I_{c} & 0\\ 0 & I_{s-c} \end{pmatrix} M' \underline{c} \\
&\ge |1-e^{i\omega}|^2  \frac{\underline{c}}{2} I_s
\end{align*}

for some real $\underline{c}$, since $|1-e^{i \omega}|^2 \le 2$
and $\frac{1}{2\pi}M'\tilde k(e^{i\omega})\Sigma \tilde k(e^{i\omega})^* M \ge \underline{c} I_s$ due to the invertibility of $\tilde k(z)$ assumed in Assumption~\ref{ass:dgp}. 
\\
The rest of the proof then follows exactly as in \cite{Palma2003}. $\qed$

\section*{Proof of Theorem~\ref{thm:cons}}
Note that $\hat \alpha(p) = \la Y_t^+ , Y_t^- \ra \la Y_t^- , Y_t^- \ra^{-1}$
is an autoregressive approximation of $Y_t^+$ by $Y_t^-$ if $p = f\tilde p$ for some integer $\tilde p$:

$$
Y_t^+ = \alpha(p) Y_t^- + U_t^+(p).
$$

\noindent \cite{Poskitt07} Theorem 5 then implies that $\| \hat \alpha(p) - \alpha(p) \|_2^2 = O(p^5 Q_T^2)$ using $(\lambda_{min}(\Gamma_p))^{-1} = O(p^{2})$ from Lemma~\ref{lem:bound}. The proof of this result in \cite{Poskitt07} can be 
easily extended to the case of general $p$ in our setting:
the argument uses $\la y_{t+j}, y_{t-k} \ra - \mE y_{t-j} y_{t-k}' = O(Q_T)$ and uniform (in $p$) bounds on the norm of $\Gamma_p$. Here $Q_T = \sqrt{(\log T)/T}$ is used. Both obviously hold also for sub-matrices. 

Clearly $\alpha(p) = \Of \Kp$ is of rank $n$ as $\Ef E_t^+$ is orthogonal to $Y_t^-$. CVA then uses a SVD of $\hWf \hat \alpha(p)  \la Y_t^- , Y_t^- \ra^{1/2}$  or equivalently the 
SVD of 

$$
\hWf \hat \alpha(p) \la Y_t^- , Y_t^- \ra \hat \alpha(p)' (\hWf)'
$$

\noindent to obtain a rank $n$ approximation where $\hWf = \la Y_t^+ , Y_t^+ \ra^{-1/2}$ (the square root denotes the Cholesky decomposition).
Due to the uniform convergence of the sample covariances we obtain $\hWf - \Wf = O(Q_T)$ for fixed $f$ since the Cholesky factorization is differentiable for positive definite matrices. 

Now $\| \alpha(p)\|_\infty = O(p)$ ($\| . \|_\infty$ denoting the row-sum norm) as can be seen, for example, from the Levinson-Whittle algorithm (see \cite{HanDei}, p. 218).  It follows that $\| \alpha(p) \|_1 = O(1)$ (column-sum norm, here equivalent to maximum entry due to finite $f$), $\alpha(p)\mE Y_{t}^- (Y_{t}^-)' = \mE Y_{t}^+ (Y_{t}^-)'$ and $\| \alpha(p) \|_2 = O(\sqrt{p})$. 
Consequently 

$$
\hat \alpha(p) \la Y_t^- , Y_t^- \ra \hat \alpha(p)' - \alpha(p) \mE Y_t^-  (Y_t^-)' \alpha(p)'
= O(p^{5/2}Q_T).
$$

The properties of the SVD then imply $\| \hOf - \Of \|_2 = O(p^{5/2}Q_T)$ which in turn leads to $\| \hKp - \Kp \|_2 = O(p^{5/2}Q_T)$: Key here is the differentiable dependence of the eigenspace to an eigenvalue on the matrix, see \cite{Chatelin93}. This applies here as $\Of$ spans the orthocomplement of the eigenspace to eigenvalue zero. The convergence for $\hOf$ then requires fixing a basis of this space which is achieved by $S_f \Of = I_n$. We then use the same normalisation for $\hOf$ such that $S_f\hOf = I_n$ to obtain  $\| \hOf - \Of \|_2 = O(p^{5/2}Q_T)$. As $\Of'\Of \ge I_s$  we have with 
$\Kp = (\Of'\Of)^{-1} \Of' \alpha(p)$ and $\hKp = (\hOf'\hOf)^{-1} \hOf' \hat \alpha(p)$ that $\|\hKp - \Kp \|_2 = O(p^{5/2}Q_T)$. 

The remainder of the proof then follows from providing error bounds for terms involving

$$
\hat x_t(p) - x_t(p)  = (\hKp - \Kp ) Y_t^-. 
$$

\noindent For example,

\begin{eqnarray*}
\la \hat x_t , \hat x_t \ra  & =   & \la \hat x_t - x_t(p), \hat x_t\ra  + \la x_t(p), \hat x_t - x_t(p) \ra  + \la x_t(p), x_t(p) \ra 
\\
 & = & (\hKp-\Kp) \la Y_{t}^- , Y_{t}^- \ra \hKp' +  \Kp \la Y_{t}^- , Y_{t}^- \ra (\hKp-\Kp)' +  \la x_t(p) , x_t(p) \ra   \\
& = & (\hKp - \Kp) \mE Y_t^- x_t(p)' + \mE x_t(p) (Y_t^-)' (\hKp - \Kp)' + 
\la x_t(p) , x_t(p) \ra  + o(p^{5/2}Q_T) \\
& = & \la x_t(p) , x_t(p) \ra  + O(p^{5/2}Q_T)
\end{eqnarray*}

\noindent where the next to last error bound follows from replacing estimates with limits. 
$\| \mE Y_t^- x_t(p)' \|_2 = O(1)$ due to the assumed stability. 
All evaluations are simple and hence omitted. 

These arguments show that uniformly for $0 < p \le H_T$ the difference between the estimates using $\hat x_t$ and using $x_t(p)$ is of order $O(p^{5/2}Q_T)$. 

Considering $\la x_t(p) , x_t(p) \ra - \mE x_t(p)x_t(p)'$ we see that 

$$
\alpha(p) (\la Y_t^- , Y_t^- \ra - \Gamma_p) \alpha(p)' = O(p^2 Q_T)
$$

\noindent since $\| \alpha(p) \|_1 = O(1)$. This holds uniformly in $p< H_T$. Similar results show that 
$\la \hat x_{t+1} , \hat x_t \ra - \mE x_{t+1}(p)x_t(p)' = O(p^{5/2}Q_T), \la y_{t} , \hat x_t \ra - \mE y_{t}x_t(p)' = O(p^{5/2}Q_T)$, $\la y_{t} , \hat x_{t+1} \ra - \mE y_{t}x_{t+1}(p)' = O(p^{5/2}Q_T)$. 
Consequently we get 

$$
\max \{ \| \hat A - A_p \|_2, \| \hat B -B_p \|_2, \| \hat C - C_p \|_2 \} = O(p^{5/2}Q_T) = o(1)
$$

for $p \le H_T$. 

To investigate $A_p - A_\circ$, for example, the difference of the second moments such as $\mE x_t(p)x_t(p)' - \mE x_t x_t'$ is essential:  
For these convergence to zero follows since $\mE \delta x_t(p) (\delta x_t(p))' \to 0$, as the approximation error converges to zero, compare Lemma~1 of \citep{Poskitt07}. 
This finishes the proof. 


%
%




\section*{Proof of Theorem~\ref{thm:asynorm}}
The proof follows the structure of the proof of asymptotic normality in \cite{BauerDeistlerScherrer}. It uses two facts:

\begin{enumerate}
    \item The covariance sequence estimators $\hat \gamma_j = \la y_t, y_{t-j} \ra, j=0,1,...,f+p$ are asymptotically normally jointly for fixed $f, p$ according to the arguments around Lemma 4.3.4 of \cite{HanDei}.
    \item The estimators of the system matrices in an appropriate overlapping form can be written as a nonlinear continuously differentiable mapping of the covariance sequence estimators. 
\end{enumerate}

The result then follows from the Delta rule. 
\\
Since 1. follows from \cite{HanDei}, we only need to investigate 2.: In this respect note that this point is used also in \cite{BauerDeistlerScherrer}. 
\\
Examine the CVA approach to construct the non-linear mapping:
\begin{itemize}
    \item Regression of $Y_t^+$ onto $Y_t^-$ is a nonlinear mapping of the covariance sequence. The estimate can be written as $\hat \beta_{f,p} = \la Y_t^+ , Y_t^- \ra \la Y_t^- , Y_t^- \ra^{-1}$. This is continuously differentiable, if $\la Y_t^- , Y_t^- \ra$ is non-singular. Since $\lambda_{min}(\Gamma_p)>0$ for fixed $p$ according to Lemma~\ref{lem:bound}, continuous differentiability holds.
    \item The second step is the calculation of the SVD involving weighting matrices. The weighting matrices are Cholesky factors of estimated second moments like $\la Y_t^- , Y_t^- \ra$. Again the weights are continuously differentiable functions of the covariance sequence. The SVD is equivalent to an eigenvalue decomposition of the squared matrix. For eigenvalue decompositions it is known that the column space is an analytic function of the matrix that is decomposed \citep[see][]{Chatelin93}. 
    \item Then $\hat \Kp = \hOf^\dagger \hat \beta_{f,p}$ is calculated. Since $\hOf$ depends continuously differentiable on the covariance sequence the same holds for $\hat \Kp$. 
    \item The remaining steps of the algorithm are regressions, that depend continuously differentiable on the second moment matrices. 
\end{itemize}

This shows the theorem. Note that the expressions in \citet{Chatelin93} even would allow for the 
derivation of expressions for the asymptotic variance. Again, given the strong bias it is questionable, if such expressions are of much value.

\end{appendices}



\bibliographystyle{elsarticle-harv} 

\begin{thebibliography}{7}
	\bibitem[{Bauer(2005)}]{Bauer2005ESTIMATINGMETHODS}
	\bibinfo{author}{Bauer, D.}, \bibinfo{year}{2005}.
	\newblock \bibinfo{title}{Estimating linear dynamical systems using subspace methods}.
	\newblock \bibinfo{journal}{Econometric Theory} \bibinfo{volume}{21}.
	pp. \bibinfo{pages}{181--211}.
	\bibitem[{Bauer(2019)}]{Bauer2019}
	\bibinfo{author}{Bauer, D.}, \bibinfo{year}{2019}.
	\newblock \bibinfo{title}{Periodic and seasonal (co-) integration in the state space framework}.
	\newblock \bibinfo{journal}{Economics Letters} \bibinfo{volume}{174}.
	pp. \bibinfo{pages}{165--168}.
	\bibitem[{Bauer and Buschmeier (2021)}]{BauerBuschmeier}
	\bibinfo{author}{Bauer, D.}, 
    \bibinfo{author}{Buschmeier, R.}, \bibinfo{year}{2021}.
	\newblock \bibinfo{title}{Asymptotic properties of estimators for seasonally cointegrated state space models obtained using the CVA subspace method}.
	\newblock \bibinfo{journal}{Entropy} \bibinfo{volume}{23}.
	\bibinfo{pages}{436}.
    	\bibitem[{Bauer, Deistler and Scherrer (1999)}]{BauerDeistlerScherrer}
	\bibinfo{author}{Bauer, D.}, 
    \bibinfo{author}{Deistler, M.}, \bibinfo{author}{Scherrer, W.}, \bibinfo{year}{1999}.
	\newblock \bibinfo{title}{Consistency and asymptotic normality of some subspace algorithms for systems without observed inputs}.
	\newblock \bibinfo{journal}{Automatica} \bibinfo{volume}{7}.
	\bibinfo{pages}{1243--1254}.
	\bibitem[{Bauer and Wagner (2021)}]{BauWag}
	\bibinfo{author}{Bauer, D.}, 
    \bibinfo{author}{Wagner, M.}, \bibinfo{year}{2012}.
	\newblock \bibinfo{title}{A state space canonical form for unit root processes}.
	\newblock \bibinfo{journal}{Econometric Theory} \bibinfo{volume}{28}.
	pp. \bibinfo{pages}{1313--1349}.
	\bibitem[{Chatelin(1993)}]{Chatelin93}
	\bibinfo{author}{Chatelin, F.}, \bibinfo{year}{1993}.
	\newblock \bibinfo{title}{{Eigenvalues of Matrices}}.
	\newblock \bibinfo{publisher}{John Wiley {\&} Sons}.
	\bibitem[{Funovits (2024)}]{Funovits}
	\bibinfo{author}{Funovits, B.}, \bibinfo{year}{2024}.
	\newblock \bibinfo{title}{{Identifiability and estimation of possibly non-invertible SVARMA Models: The normalised canonical WHF parametrisation}}.
	\newblock \bibinfo{journal}{Journal of Econometrics}
	\bibinfo{volume}{241}, \bibinfo{pages}{105766}.
	\bibitem[{Hannan and Deistler(1988)}]{HanDei}
	\bibinfo{author}{Hannan, E.J.}, \bibinfo{author}{Deistler, M.},
	\bibinfo{year}{1988}.
	\newblock \bibinfo{title}{{The Statistical Theory of Linear Systems}}.
	\newblock \bibinfo{publisher}{John Wiley}, \bibinfo{address}{New York}.
	\bibitem[{Larimore(1983)}]{Larimore1983}
	\bibinfo{author}{Larimore, W.E.}, \bibinfo{year}{1983}.
	\newblock \bibinfo{title}{{System Identification, reduced order filters and
			modeling via canonical variate analysis}}, \bibinfo{address}{Piscataway, NJ}.
	pp. \bibinfo{pages}{445--451}.
	\bibitem[{Palma and Bondon(2003)}]{Palma2003}
	\bibinfo{author}{Palma, W.}, \bibinfo{author}{Bondon, P.},
	\bibinfo{year}{2003}.
	\newblock \bibinfo{title}{{On the Eigenstructure of Generalized Fractional
			Processes}}.
	\newblock \bibinfo{journal}{Statistics and Probability Letters}
	\bibinfo{volume}{65}, \bibinfo{pages}{93--101}.
	\bibitem[{P\"otscher (1991)}]{Poetscher}
	\bibinfo{author}{P\"otscher, B.}, \bibinfo{year}{1991}.
	\newblock \bibinfo{title}{Noninvertibility and pseudo-maximum likelihood estimation of misspecified ARMA models}.
	\newblock \bibinfo{journal}{Econometric Theory}
	\bibinfo{volume}{7}, \bibinfo{pages}{435--449}.  
	\bibitem[{Poskitt(2006)}]{Poskitt07}
	\bibinfo{author}{Poskitt, D.S.}, \bibinfo{year}{2006}.
	\newblock \bibinfo{title}{{Autoregressive approximation in nonstandard
			situations: the fractionally integrated and non-nivertible case}}.
	\newblock \bibinfo{journal}{Annals of Institute of Statistical Mathematics}
	\bibinfo{volume}{59}, \bibinfo{pages}{697--725}.
	
\end{thebibliography}

\end{document}